# Implicit Large Eddy Simulation of Cavitation in Micro Channel Flows

S. Hickel * ,  M. Mihatsch ,  S.J. Schmidt

Institute of Aerodynamics and Fluid Mechanics, Technische Universität München
85747 Garching, Germany.

* email: sh@tum.de

**ABSTRACT**

We present a numerical method for Large Eddy Simulations (LES) of compressible two-phase flows. The method is validated for the flow in a micro channel with a step-like restriction. This setup is representative for typical cavitating multi-phase flows in fuel injectors and follows an experimental study of Iben et al. [1]. While a diesel-like test fuel was used in the experiment, we solve the compressible Navier-Stokes equations with a barotropic equation of state for water and vapor and a simple phase-change model based on equilibrium assumptions. Our LES resolve all wave dynamics in the compressible fluid and the turbulence production in shear layers.

## 1. INTRODUCTION

Turbulence modeling and the numerical discretization of the Navier-Stokes equations are strongly coupled in Large Eddy Simulations (LES). Since subgrid-scale (SGS) models generally operate on scales that are only marginally resolved by the underlying numerical method, the truncation error of common approximations for the convective terms can out-weigh the effect of even physically sound models. This interference is especially severe in LES of cavitating flows, as numerical errors of robust discretizations increase notably at discontinuities such as phase boundaries and shock waves that are generated by vapor-bubble collapses.

A different approach is to exploit this link by developing discretization methods from subgrid-scale models, or vice versa. Approaches where SGS models and numerical discretizations are fully merged are called implicit LES (ILES). SGS effects are modeled explicitly if the underlying conservation law is modified and subsequently discretized. The filtering concept of Leonard [2] is commonly employed for deriving explicit SGS models without reference to a computational grid and without taking into account a discretization scheme. As implicit modeling we denote the situation when the unmodified conservation law is discretized and the numerical truncation error acts as an SGS model. Since this SGS model is implicitly contained within the discretization scheme, an explicit computation of model terms becomes unnecessary.

Most previous approaches to implicit SGS modeling have relied on the application of pre-existing discretization schemes to fluid-flow turbulence. Consequently, methods with suitable implicit SGS models have usually been found by trial and error, which has led to the common belief that an implicit subgrid-scale model is merely inferred by the choice of discretization. Comparative studies have shown that stabilizing under-resolved simulations by upwind or non-oscillatory schemes is insufficient for accurately representing SGS turbulence. For example, Honein and Moin [3] found that traditional ILES required differently tuned parameters to predict the correct decay rates of different quantities. Employing implicit LES for prediction requires numerical methods that are specially designed, optimized and validated for the particular differential equation to be solved. A full coupling of the SGS model and the discretization scheme cannot be achieved without incorporating physical reasoning into the design of the implicit SGS model.

Implicit SGS modeling requires procedures for design, analysis, and optimization of nonlinear discretization schemes. We have proposed such a systematic framework that resulted in the adaptive local deconvolution method (ALDM), see Refs. [4,5]. ALDM is a nonlinear finite volume scheme based on a solution adaptive deconvolution operator and a numerical flux function. Free parameters inherent to the discretization allow to control the truncation error and have been calibrated in such a way that the truncation error acts as a physically motivated SGS model. The compressible version of ALDM, see Ref. [5], is asymptotically consistent with incompressible turbulence theory, which is essential for cavitating low Mach number flows. Yet, the numerical discretization is robust enough to survive strong shock waves [6].

In this paper, we present the first implicit LES of a compressible cavitating two-phase flow. The thermodynamic model and the numerical method are derived in Sections 2 and 3, respectively. A validation of the novel method is presented in Section 4. As the test case we chose a micro channel with a step-like restriction, which is representative for typical cavitating-flow situations in fuel injectors. This setup follows closely an experimental study of Iben et al. [1]. While a diesel-like fuel with confidential properties was used in the experiment, we solve the compressible Navier-Stokes equations with a barotropic equation of state for vapor and water and a simple phase-change model.

## 2. MATHEMATICAL AND PHYSICAL MODEL

### 2.1 Governing equations

The governing equations in our simulations are the fully compressible three-dimensional Navier-Stokes equations written in integral form

$$\partial_t \overline{U} = \frac{1}{V} \oint_{\partial V} (C(U) + S(U)) \cdot dA \quad (1)$$

with appropriate initial and boundary conditions. The solution vector containing the volume-averaged conserved variables is

$$\overline{U} = \frac{1}{V} \int_V U dV, \quad (2)$$

where the vector $U = [\rho, \rho u_1, \rho u_2, \rho u_3, \rho E]$ represents density, momentum and total energy. The overbar denotes the volume average for an arbitrary control volume $V$ with surface $\partial V$. Note that we distinguish between the convective flux

$$C_i(U) = [u_i \rho, u_i \rho u_1, u_i \rho u_2, u_i \rho u_3, u_i \rho E] \quad (3)$$

and the flux due to surface stresses

$$S(U) = [0, \delta_{i1} p - \tau_{i1}, \delta_{i2} p - \tau_{i2}, \delta_{i3} p - \tau_{i3}, u_k (\delta_{ik} p - \tau_{ik}) + q_i], \quad (4)$$

where $u_i$ is the velocity vector, $\tau_{ij}$ denotes the viscous stress tensor

$$\tau_{ij} = \mu \left( \partial_j u_i + \partial_i u_j - \frac{2}{3} \delta_{ij} \partial_k u_k \right), \quad (5)$$

and the heat flux is

$$q_i = -\kappa \partial_i T. \quad (6)$$

The numerical reconstruction of $C(U)$ and $S(U)$ at the surface $\partial V$ will be discussed in Section 3.

As we consider cavitating flows with phase change, the fluid within the volume $V$ can be in liquid or gaseous state. For later convenience we define the vapor volume fraction

$$\alpha = \frac{V_{vap}}{V}, \quad (7)$$

where $V_{vap}$ denotes the part of the volume $V$ that is occupied by vapor. Analogously, $V_{liq} = V - V_{vap}$ is volume of fluid in liquid state. The Navier-Stokes equations (1) are closed by constitutive relations for the pressure $\overline{p}$ and viscosity $\overline{\mu}$ of the volume-averaged, i.e. homogenized, fluid.

### 2.2 Thermodynamic model

Our cavitation model is based on a barotropic equation of state for water and water-vapor mixtures, i.e., the pressure and vapor volume fraction are pure functions of the mean density

$$\overline{\rho} = \alpha \rho_{vap} + (1 - \alpha) \rho_{liq}. \quad (8)$$

Phase change is supposed to be infinitely fast, isentropic and in mechanical equilibrium. The shape of the liquid-vapor interface is not reconstructed and the effect of surface tension on the vapor pressure is neglected. With these assumptions, the densities of liquid and vapor are $\rho_{liq} = \rho_{sat,liq}$ and $\rho_{vap} = \rho_{sat,vap}$, i.e., we can compute the vapor volume fraction from

$$\alpha = \frac{V_{vap}}{V} = \begin{cases} 0 & , \overline{\rho} \geq \rho_{sat,liq} \\ \frac{\rho_{sat,liq} - \overline{\rho}}{\rho_{sat,liq} - \rho_{sat,vap}} & , \overline{\rho} < \rho_{sat,liq} \end{cases}, \quad (9)$$

where $\rho_{sat,liq}$ and $\rho_{sat,vap}$ are saturation densities of pure water and pure vapor, respectively. Liquid water ($\alpha=0$) is modeled with high accuracy by a modified Tait equation of state

$$\overline{p} = (p_{sat} + B) \cdot \left( \frac{\overline{\rho}}{\rho_{sat,liq}} \right)^N - B \quad , \alpha = 0 \quad (10)$$

where $N = 7.1$ and $B = 3.06 \cdot 10^8$ Pa are fitted constants. Water starts to evaporate if the pressure drops below the saturation pressure $p_{sat}$. Assuming phase change along an isentropic equilibrium path, we obtain the following equation for the equilibrium speed of sound

$$\frac{1}{\overline{\rho} c_{eq}^2} = \left( \frac{\alpha}{\rho_{vap} c_{vap}^2} + \frac{1-\alpha}{\rho_{liq} c_{liq}^2} \right) + \left( \frac{1}{\rho_{vap}} + \frac{1}{\rho_{liq}} \right)^2 \frac{T \left( \alpha \rho_{vap} c_{p,vap} + (1-\alpha) \rho_{liq} c_{p,liq} \right)}{L}. \quad (11)$$

$L$ is the latent heat at the temperature $T$ and $c_{p,liq}$ and $c_{p,vap}$ are the specific heats at constant pressure of the liquid and of the vapor. As $\rho_{liq} >> \rho_{vap}$, the equilibrium speed of sound in the two-phase regime, i.e. for $0<\alpha<1$, is

$$c_{eq} = \frac{L \rho_{vap}}{\sqrt{c_{p,liq} T}} \frac{1}{\overline{\rho}} + HOT. \quad (12)$$

By integration of the square of Eq. (12) we obtain the equilibrium pressure

$$\overline{p} = p_{sat} + C \left( \frac{1}{\rho_{sat,liq}} - \frac{1}{\overline{\rho}} \right), \quad 0 < \alpha < 1. \quad (13)$$

The parameter $C$, as well as the pressure and density at the saturation point generally depend on temperature. This temperature dependence is neglected in our isentropic model. For the chosen reference temperature of $T = T_{ref} = 293.15$ K follow $p_{sat} = 2340$ N/m$^2$, $\rho_{sat,liq} = 998.1618$ kg/m$^3$, $\rho_{sat,vap} = 0.01731$ kg/m$^3$, and $C = 1468.54$ Pa kg/m$^3$.

The dynamics of wall-bounded flows, such as the flow through micro channels, is strongly influenced by viscous stresses. As the dynamic viscosities of liquid water and vapor differ by several orders of magnitude, the chosen model for the viscosity of the homogenized two-phase fluid is of particular importance. Frequently [7-8, e.g.], a straightforward approach

$$\overline{\mu} = \alpha \mu_{vap} + (1 - \alpha) \mu_{liq} \quad (14)$$

is used, which assumes unlimited deformability. The surface tension at the phase boundaries, however, can increase the resistance against deformation substantially. Fully immersed small vapor bubbles behave approximately like rigid particles. Einstein [9] found

$$\bar{\mu} = (1 + \frac{5}{2}\alpha)\mu_{liq} \qquad (15)$$

for the effective viscosity of a suspension of a large number of small particles in a liquid. We follow Ref. [10] and combine Eq. (14) and Eq. (15) to obtain the final form of the volume averaged viscosity of the cavitating fluid

$$\bar{\mu} = (1-\alpha)(1 + \frac{5}{2}\alpha)\mu_{liq} + \alpha\,\mu_{vap} \qquad (16)$$

with $\mu_{liq} = 1.002\ 10^{-3}$ Pa s and $\mu_{vap} = 9.727\ 10^{-6}$ Pa s.

## 3. NUMERICAL METHOD

### 3.1 Implicit LES

Our flow solver INCA is a finite volume method for LES. Deriving suitable numerical models for LES of compressible flows is particularly challenging because the unresolved subgrid scales in LES represent very different flow phenomena. On one hand, SGS can originate from turbulence. On the other hand, there are unique (resolution invariant) discontinuities, such as shock waves and material interfaces, which are very distinct from the smooth variations produced by turbulence and require different modeling approaches.

For phenomena involving mechanisms that may invalidate underlying assumptions of classical turbulence models, it is necessary to rely on more elaborate approaches such as implicit LES. Implicit LES modeling involves a direct coupling between the numerical scheme and the SGS model. For brevity of notation, the following summary of this concept is given for the 1-D case and a generic non-linear transport equation

$$\partial_t u + \partial_x F(u) = 0 \qquad (17)$$

Following Leonard [2] the discretized equation is obtained by convolution with a homogeneous filter kernel $G$ and subsequent discretization

$$\partial_t \bar{u}_N + G * \partial_x F_N(u_N) = -G * \partial_x \tau_{SGS} \qquad (18)$$

where the overbar denotes the filtering

$$\bar{u} = G * u \qquad (19)$$

and the subscript $N$ indicates grid functions obtained by projecting continuous functions onto the grid $x_N = \{\ x_j\ \}$. The subgrid-stress tensor

$$\tau_{SGS} = F(u) - F_N(u_N) \qquad (20)$$

originates from the grid projection of non-linear terms and has to be modeled in order to close Eq. (18). With Leonard's approach the filtering of the continuous system is considered as predominant approximation where the numerical error in solving this continuous system is supposed to be negligibly small. Consequently, explicit SGS models are usually derived without reference to a computational grid and without taking into account any discretization scheme.

The unfiltered, i.e., continuous, solution $u$ is unknown in LES. However, accurate approximations $\tilde{u}_N$ of the discrete grid function $u_N$ can be reconstructed from $\bar{u}_N$ by regularized deconvolution. Hence, the solution obtained with the discrete operators does not satisfy Eq. (18), but rather a modified differential equation

$$\partial_t \bar{u}_N + G * \partial_x F_N(u_N) = \varepsilon_N - \tilde{G} * \tilde{\partial}_x \tilde{\tau}_{SGS} \qquad (21)$$

Solved numerically, the discrete approximation of the SGS model interferes with the truncation error of the underlying discretization scheme

$$\varepsilon_N = G * \partial_x F_N(u_N) - \tilde{G} * \tilde{\partial}_x \tilde{F}_N(\tilde{u}_N), \qquad (22)$$

where a tilde indicates the respective numerical approximation. This interference of model and discretization is exploited in implicit LES. Particularly, an explicit SGS model is resembled if the filtered divergence of the model SGS tensor is approximated by

$$\varepsilon_N \approx -G * \partial_x \tau_{SGS} \qquad (23)$$

so that no model terms have to be computed explicitly during time advancement.

### 3.2 Adaptive Local Deconvolution Method (ALDM)

With ALDM numerical discretization and SGS modeling are merged entirely. The discrete system for evolving a grid-function approximation is considered directly as a truncated representation of the unmodified continuous system, Eqs. (1) and (17). A suitable environment for the discretization design is provided by a finite-volume method that works directly on the integral form of the Navier-Stokes equations (1). Although filtering is not performed explicitly, we can use the filter formulation of Leonard as an analytical tool when designing and analyzing the discrete operators. A finite-volume discretization corresponds to the evaluation of Eq. (18) with a top-hat filter

$$G(x, V_j) = \begin{cases} 1/V_j & , x \subset V_j \\ 0 & , \text{else} \end{cases}, \qquad (24)$$

where $V_j$ denotes the cell volume (3-D) or the grid spacing (1-D) of the cell $j$ of the computational grid. Note the equivalence of Eqs. (19) with (24) and Eq. (2).

The numerical building blocks of finite-volume methods are a reconstruction of the unfiltered solution at cell faces, a numerical flux function that works on the reconstructed solution, and a numerical integration scheme to compute the face-averaged flux. The averaging and reconstruction steps involved in finite-volume discretizations are related to the filtering and deconvolution that are well known in explicit SGS modeling. Following the concept of Schumann [11], the subgrid-stress tensor has to account for the limited numerical accuracy of the flux reconstruction. With respect to implicit LES, an advantageous aspect is the fact that the numerical truncation error of the reconstruction readily appears as a divergence of a tensor, as required for physically motivated implicit modeling by Eq. (23).

With ALDM, a local reconstruction of the unfiltered solution is obtained from a solution-adaptive combination

$$\tilde{u}^{\mp}(x_{j\pm 1/2}) = \sum_{k=1}^{K}\sum_{r=0}^{k-1}\omega_{kr}^{\mp}(\gamma_{kr},\bar{u}_N)\tilde{g}_{kr}^{\mp}(x_{j\pm 1/2}) \qquad (25)$$

of Harten-type deconvolution polynomials

$$\tilde{g}_{kr}^{\mp}(x_{j\pm 1/2}) = \sum_{l=0}^{k-1} c_{krl}^{\mp}(x_N)\bar{u}_{j-r+l}, \qquad (26)$$

where half-integer indices denote reconstructions at the cell faces. The grid-dependent coefficients $c_{krl}^{\mp}$ are chosen such that

$$\tilde{g}_{kr}^{\mp}(x_{j\pm 1/2}) = u(x_{j\pm 1/2}) + O(h^k) \qquad (27)$$

on a grid with spacing $h$. Adaptivity of the deconvolution operator is achieved by dynamically weighing the respective contributions by $\omega_{kr}^{\mp}(\gamma_{kr},\bar{u}_N)$, where $\gamma_{kr}$ are free model parameters. Instead of maximizing the order of accuracy, deconvolution is regularized by limiting the degree $k$ of local approximation polynomials to $k \leq K$ and by permitting all polynomials of degree $1 \leq k \leq K=3$ to contribute to the approximately deconvolved solution. The computational cost of a multi-dimensional finite-volume scheme strongly depends on the implementation of the reconstruction step. An efficient method for the 3-D reconstruction scheme of ALDM is given in Ref. [12]. This simplified adaptive local deconvolution (SALD) method reduces the amount of computational operations without affecting the quality of the LES results. The SALD implementation is used for all computations in this paper.

### 3.3 Mimetic flux function for barotropic fluids

A flux function suitable for implicit LES of cavitating flows has to meet the following requirements:

1. As simple and efficient as possible to facilitate computations.

2. Preservation of symmetries of the Navier-Stokes equations and asymptotic consistency with incompressible turbulence theory for low-Mach-number flows.

3. Robust and able to capture discontinuities, such as phase boundaries and shock waves.

Standard flux functions for inviscid flows, such as the HLL method of Harten, Lax and van Leer [17], e.g., satisfy point 3 by an approximate solution of the Riemann problem $(\tilde{u}^-,\tilde{u}^+)$. Representing a discontinuity, this solution is however inconsistent with low-Mach-number turbulence, c.f. point 2, in which SGS originate from smooth variations. Thus it is not surprising that shock-capturing Euler schemes strongly over-estimate the SGS dissipation of shock-free turbulence [18]. In the sense of Kovasznay [19], compressible LES must correctly model vorticity, entropy and acoustic modes, while the first mode alone is sufficient in incompressible LES. It would stand to reason to use a flux function that works on the reconstructed characteristic variables. The required local decomposition in vorticity, entropy and acoustic modes, however, increases the computational costs by an order of magnitude, c.f. point 1, and can lead to stability problems if the material properties, such as the speed of sound, are discontinuous functions of the conserved variables. Simple flux upwinding is computationally efficient and can provide sufficient dissipation at discontinuities; however, implicit LES requires a Galilean invariant and more elaborate approach with the possibility of controlling the numerical dissipation dynamically.

In the ALDM framework, secondary regularization is provided by a suitable consistent numerical flux function with the general form

$$\tilde{F}_{j\pm 1/2} = F\left(\frac{\tilde{u}^+_{j\pm 1/2} + \tilde{u}^-_{j\pm 1/2}}{2}\right) - R(\sigma,\tilde{u}_N^{\pm},\bar{u}_N). \qquad (28)$$

This numerical flux consists of two parts. The first term corresponds to the physical Navier-Stokes flux. For maximum order of consistency it is computed from the mean of both reconstructions of the unfiltered solution at the considered cell face. Note that this results in a skew-symmetric like discretization with favorable aliasing behavior. The difference between both interpolants is exploited in the regularization term

$$R_{j\pm 1/2} = D(\sigma,\tilde{u}_N^{\pm},\bar{u}_N)(\tilde{u}^+_{j\pm 1/2} - \tilde{u}^-_{j\pm 1/2}) \qquad (29)$$

working on the reconstruction error $\tilde{u}^+_{j\pm 1/2} - \tilde{u}^-_{j\pm 1/2}$. The dissipation matrix $D$ can be any non-negative, shift-invariant functional of $\bar{u}_N$ and $\tilde{u}_N$, which needs to be defined specifically for the particular differential equation.

The thermodynamic equilibrium model derived in Sec. 2.2 is based on a barotropic equation of state for the homogenized single- and two-phase fluid. In the following, we therefore only consider the mass and momentum flux. The local adaptive reconstruction scheme of ALDM is applied to the primitive variables density, velocity and pressure. Gradients in the stress tensor $S$ are approximated by linear second order schemes.

We observe that the convective flux for a barotropic fluid can be written as

$$C_i(U) = u_i \rho [1, u_1, u_2, u_3] \qquad (30)$$

and propose that also the numerical flux should satisfy this property. The numerical approximation of the mass density flux at an arbitrary cell face is

$$\tilde{C}^{\rho} = \tilde{u}^C \frac{\tilde{\rho}^+ + \tilde{\rho}^-}{2} - D^{\rho}(\tilde{\rho}^+ - \tilde{\rho}^-). \qquad (31)$$

The dissipation matrix $D$ and the transport velocity $\tilde{u}^C$ will be defined later. The mimetic numerical flux for the momentum equations fulfilling property (30) is

$$\tilde{C}^{\rho u} = \tilde{C}^{\rho} \frac{\tilde{u}^+ + \tilde{u}^-}{2} - D^{\rho u} \frac{\tilde{\rho}^+ + \tilde{\rho}^-}{2}(\tilde{u}^+ - \tilde{u}^-). \qquad (32)$$

For defining the transport velocity $\tilde{u}^C$, we follow the geometric argument of Harten, Lax and van Leer [17] and propose the following modification of the contact wave speed of the HLL approximate Riemann solver

$$\tilde{u}^C = \frac{\tilde{u}^+ + \tilde{u}^-}{2} - \frac{\tilde{p}^+ - \tilde{p}^-}{\tilde{\rho}^+(S_R - \tilde{u}^+) - \tilde{\rho}^-(S_L - \tilde{u}^-)}, \qquad (33)$$

where SR and SL denote conservative estimates of the fastest right and left going wave speeds, which we compute from

$$\begin{aligned} S_R &= \max(\tilde{u}^+, \tilde{u}^-) + c_{liq} \\ S_L &= \min(\tilde{u}^+, \tilde{u}^-) - c_{liq} \end{aligned}. \qquad (34)$$

In order to ensure asymptotic consistency of the flux with ALDM for incompressible turbulence and passive scalar mixing, Ref. [4,13], in the limit of small Mach numbers $Ma = u/c \to 0$, we define the dissipation matrix as

$$D_i = \begin{bmatrix} \sigma^\rho |\delta\bar{u}_i| \\ \sigma^{\rho u} |\delta\bar{u}_1| \\ \sigma^{\rho u} |\delta\bar{u}_2| \\ \sigma^{\rho u} |\delta\bar{u}_3| \end{bmatrix}. \qquad (35)$$

where $\sigma^\rho$ and $\sigma^{\rho u}$ are free model parameter and $\delta\bar{u}$ denotes the difference of volume-averaged velocity of the two adjacent cells.

The free parameters $\{\gamma,\sigma\}$ in the solution-adaptive stencil-selection scheme and the numerical flux function can be used to adjust the spatial truncation error of the discretization. For implicit SGS modeling, the values of these free parameters are chosen in such a way that the truncation error $\varepsilon_N$ of the discretization method acts as a physically motivated SGS model. We use optimized parameter values that were determined in Refs. [4,13] by minimizing a cost function that measures the difference between the effective spectral numerical viscosity of ALDM and the eddy viscosity from Eddy-Damped Quasi-Normal Markovian (EDQNM) theory for isotropic turbulence. The resulting method represents a full merger of numerical discretization and SGS model. Incorporating the essential elements of LES, filtering and deconvolution, the implicit model of ALDM combines a tensor-dissipation regularization with a generalized scale-similarity approach. Explicit deconvolution-type SGS models have so far been limited to linear deconvolution. ALDM extends, by employing methods that are well established for essentially non-oscillatory finite-volume discretizations, the concept of approximate deconvolution to the solution-adaptive non-linear case. Finally, we note that the solution-adaptive stencil selection and the conservative flux formulation at the core of the present method allow for the capturing of shock waves.

The spectral energy transfer at shock waves differs considerably from the mechanisms in unresolved turbulence. These differences have to be taken into account in the SGS modeling strategy. Unresolved turbulence can be modeled with a numerical viscosity proportional to a velocity gradient and the square of a characteristic length scale. Propagating discontinuities, which represent unique SGS, are usually modeled with a numerical viscosity that is proportional to the wave propagation speed. In order to capture shock waves with ALDM, we detect discontinuities by the sensor functional

$$f_S = \begin{cases} 1, & \frac{|div(u)|}{|div(u)| + \|rot(u)\|} > 0.95 \\ 0, & \text{else} \end{cases} \qquad (36)$$

and locally add a second term to the dissipation matrix

$$D_i = \begin{bmatrix} \sigma^\rho |\delta\bar{u}_i| \\ \sigma^{\rho u} |\delta\bar{u}_1| \\ \sigma^{\rho u} |\delta\bar{u}_2| \\ \sigma^{\rho u} |\delta\bar{u}_3| \end{bmatrix} + f_S \frac{1}{2}\left(|\tilde{u}^C| + |\delta\bar{u}_i|\right) \begin{bmatrix} 1 \\ 1 \\ 1 \\ 1 \end{bmatrix}. \qquad (37)$$

Note that consistent behavior in the limit of low Mach numbers is maintained by using $|\tilde{u}^C| + |\delta\bar{u}_i|$ rather than $|\tilde{u}^C| + c$.

### 3.4 Conservative immersed boundary method

The generation of suitable grids for LES of complex flows can be time-consuming and difficult. Contradictory requirements, such as adequate local resolution and minimum number of grid points, can deteriorate the grid quality and adversely affect accuracy and numerical convergence properties. Cartesian grids facilitate automatic grid generation and adaptive local grid refinement by dyadic sub-partitioning, which can be easily accounted for in implicit SGS modeling. Cartesian grids also imply fewer computational operations per grid point than body-fitted or unstructured grids. On the other hand, geometric boundaries do not necessarily coincide with grid lines, so that boundary conditions have to be applied at the subcell level.

We use a conservative immersed boundary method for representing sharp interfaces between a fluid and a rigid body on Cartesian grids, see Refs. [14,15]. A level-set technique is used for describing the interface geometry. The level-set function is the signed distance between each point in the computational domain and the immersed interface, which is positive within and negative outside of the fluid domain. The zero contour is the interface between the fluid and the obstacle.

The intersection of the obstacle with the Cartesian grid produces a set of cells that are cut by the interface. The underlying finite-volume discretization is modified locally in these cut cells in such a way that it accounts only for the fluidic part of a cell. The viscous stresses at the fluid-solid interface are approximated by linear differencing schemes. The interface pressure can be obtained by solving the one-sided (symmetric) Riemann problem in the interface-normal direction, e.g.

$$p^* = p_L + \rho_L(u_L - S_L)(u_L - S_C), \qquad (38)$$

where the contact wave velocity is $S_C = 0$ and the left going characteristic is $S_L = \min(u_L, S_C) - c_{eq}$ for non-moving walls.

Because Eq. (38) is inconsistent in the limit $Ma = u/c \to 0$, a Neumann boundary condition

$$p^* = p_L \qquad (39)$$

is imposed in the well resolved boundary layers of the present LES, where the local Mach number is low in cut cells.

As we operate on fluxes only, this cut-cell finite-volume method maintains accuracy and ensures mass and momentum conservation. Required interface normals, face apertures and fluid-volume fractions of cut cells can be computed efficiently from the level-set function. Discrete conservation and a sharp representation of the fluid-solid interface render our method particularly suitable for LES of turbulent flows.

### 3.5 Time integration

For time integration, the conditionally stable 3$^{rd}$ order Runge-Kutta method of Gottlieb and Shu [16] is used. This time-discretization scheme is total-variation diminishing (TVD) for CFL≤1, provided the underlying spatial discretization is TVD, whereas the linear stability bound is larger. We found for ALDM stable time advancement up to the linear bound CFL=1.7. As cut cell methods can generate cells with a very small fluid volume fraction, a special treatment of such cells is necessary when excessively small time steps according to the CFL stability criterion are to be avoided. For increasing the computational efficiency, we use a conservative mixing procedure, where the conservatives of small cells are mixed with larger neighbors [14,15]. During our simulations, the time step size is adjusted dynamically according to CFL=0.5 based on the full cell size.

## 4. COMPUTATIONAL SETUP AND RESULTS

### 4.1 Setup

The new ILES methodology is applied to the compressible turbulent flow in a micro channel with a step-like restriction. This forward-facing step geometry, as shown in Fig. 1, was investigated experimentally by Iben *et al.* [1] and is representative for typical cavitating-flow situations in fuel injectors.

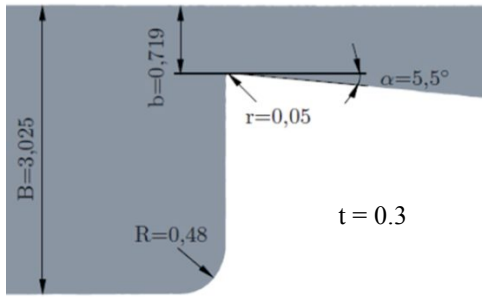

**Fig. 1:** Geometry of forward facing step flow.

The inlet is located 6 mm upstream of the restriction and has a cross section of 3.0x0.3 mm$^2$. A laminar channel-flow profile with a bulk velocity of 12 m/s and a mass flow of 10.8 g/s is imposed at the inflow. A static pressure of 50 - 60 bar is imposed at the outlet. The flow is accelerated at the restriction (0.7x0.3 mm$^2$) and the pressure drops close to saturation pressure. A shear layer, which is subject to Kelvin-Helmholtz instability, is formed on top of the forward facing step. Cavitation is first observed in the vortex cores of this shear layer. Fig. 2 shows a snapshot of the instantaneous cavitation distribution in the experiment [1].

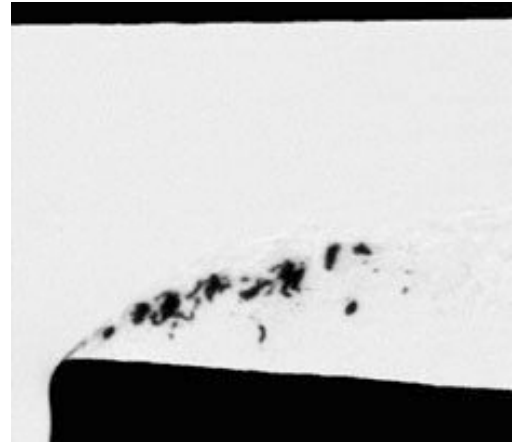

**Fig. 2:** Instantaneous cavitation distribution in the experiment of Iben *et al.* [1].

The flow geometry is represented by our flow solver INCA through a conservative cut-cell method on an adaptive Cartesian grid. A suitable initial solution for our LES is generated through a grid sequencing technique. That is, we start the simulation on an initially very coarse grid that is gradually refined during the simulation as the solution approaches a developed turbulent state. The final grid is shown in Fig. 3. It consists of 11x10$^6$ cells and has a resolution of 2x10$^{-6}$m in the region of interest. The fine grid resolution and the conditionally stable time integration method imply a time step size of 2.3x10$^{-10}$s. Several million time steps were required to resolve the integral time scales of the flow. The present LES was only feasible because the grid is rapidly coarsened towards the inflow and outflow. Fig. 3 shows also a large reservoir that is attached at the right side in order to reduce reflections of acoustic waves at the fixed-pressure outflow boundary condition.

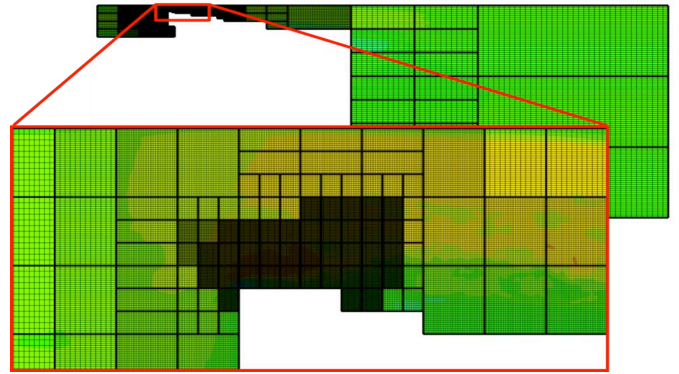

**Fig. 3:** Adaptive Cartesian computational grid. The channel geometry, which is represented by a conservative cut-cell method with sub-cell accuracy, is not shown.

## 4.2 Results

The initial data for a developed turbulent flow was generated with an outlet pressure of $p_{out}$=60 bar. The backpressure is then gradually reduced to 57, 54, and 50 bar. Fig 4 shows the evolution of the mean vapor volume fraction (based on the volume of the entire computational domain). Cavitation is first observed in the vortex cores of the shear layer for $p_{out}$=57 bar. Reduced outlet pressures of $p_{out}$=54 bar and eventually $p_{out}$=50 bar lead to a higher cavitation probability, as can be seen in the insert of Fig. 4.

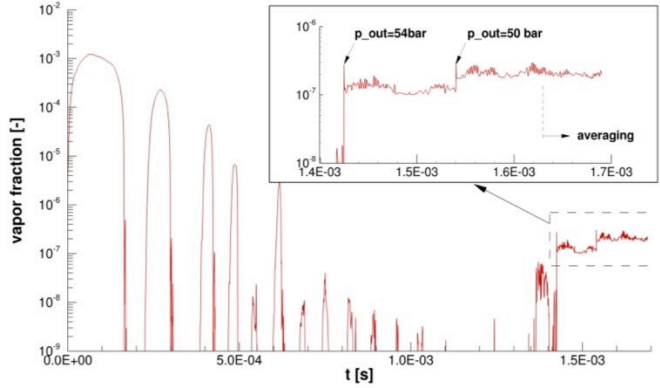

**Fig. 4:** Evolution of the mean vapor volume fraction.

We visualize the vapor regions and corresponding coherent turbulence structures at the onset of cavitation ($p_{out}$=57 bar) in Fig. 5. We observe large incoming corner vortices, which lead to flow separation in the corners between the channel sidewalls and the restriction. The displacement by these separation bubbles results in further acceleration of the bulk flow. The flow separation renders the flow 3-D. 2-D Kelvin-Helmholtz rollers are not visible in contour plots of the $\lambda_2$ vortex-identification criterion; rather hairpin vortices are observed in the unstable shear layer. Vapor regions are clearly correlated with the cores of these vortices. This finding is consistent with the experimental observations, c.f. Fig. 2.

Figures 6 and 7 show instantaneous velocity, density and pressure fields in the center plane above the restriction for $p_{out}$=50 bar. We clearly see the shear layer and corresponding turbulent structures. A statistical analysis has been conducted for the final backpressure level of 50 bar. 70.000 statistical samples (gathered every 10 time steps) are averaged in time and in the spanwise direction. The spanwise averaging improves convergence and comparability with the experiment, where possible observations through side windows represent spanwise optical averages. The mean velocity and mean void fraction is shown in Fig. 8, Reynolds stresses are shown in Fig. 9.

The mean vapor fraction, lower part of Fig. 8, is in reasonable agreement with experimental data in Fig. 5 of Ref. [1]. We cannot expect perfect agreement, because our LES were performed with the material properties of water, whereas a diesel-like fuel with confidential properties was used in the experiment. The values of our inlet mass flux and outlet pressure therefore constitute an educated guess rather than exact data. We should notice, however, that the onset of cavitation appears to happen slightly too early, which can be attributed to the thermodynamic equilibrium model. We also notice that the grid coarsening in our LES imposes a fixed bound for the location of the rear end of the cavitating two-phase region, see e.g. the lower part of Fig. 7. A simulation with an extended fine-grid zone is therefore planed for the future.

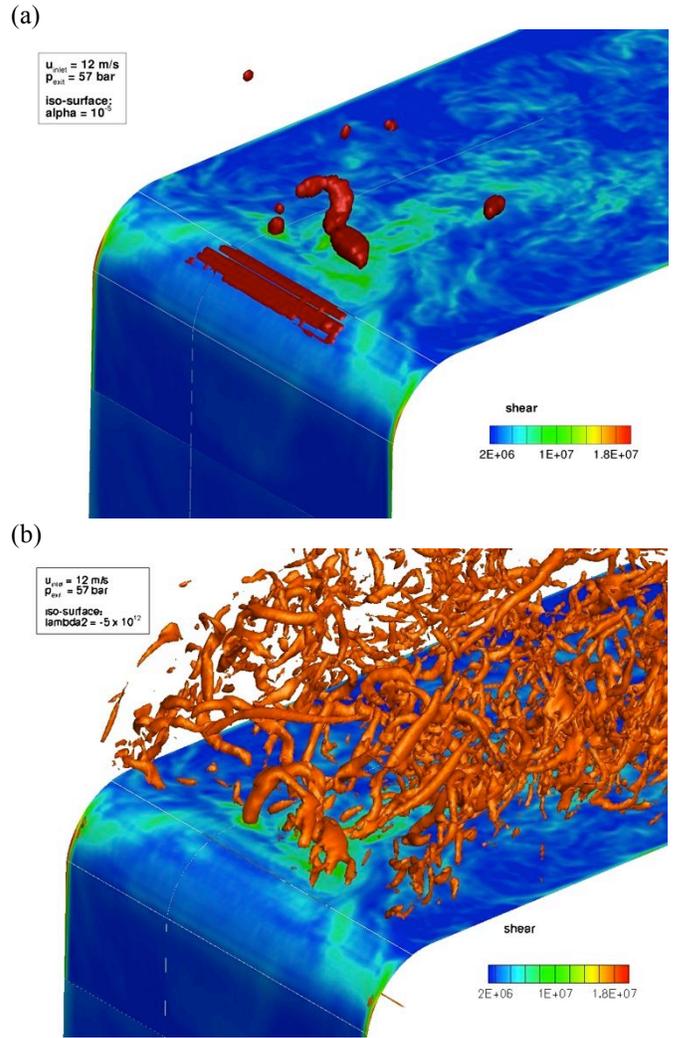

**Fig. 5:** LES of cavitating forward-facing step flow in a micro channel. (a) Shear stress at the wall and iso-surface of the vapor volume fraction visualizing cavitation. (b) Coherent turbulence structures visualized by the $\lambda_2$ criterion.

## 5. CONCLUSIONS

We have presented a numerical method for implicit LES of compressible two-phase flows. The novel method is applied and validated for the flow in a micro channel with a step-like restriction. Our LES resolve all wave dynamics in the compressible fluid and the turbulence production in shear layers. The numerical results are in good agreement with available experimental data. This work represents, to our knowledge, the first successful LES of a cavitating two-phase flow with a compressible fluid model.

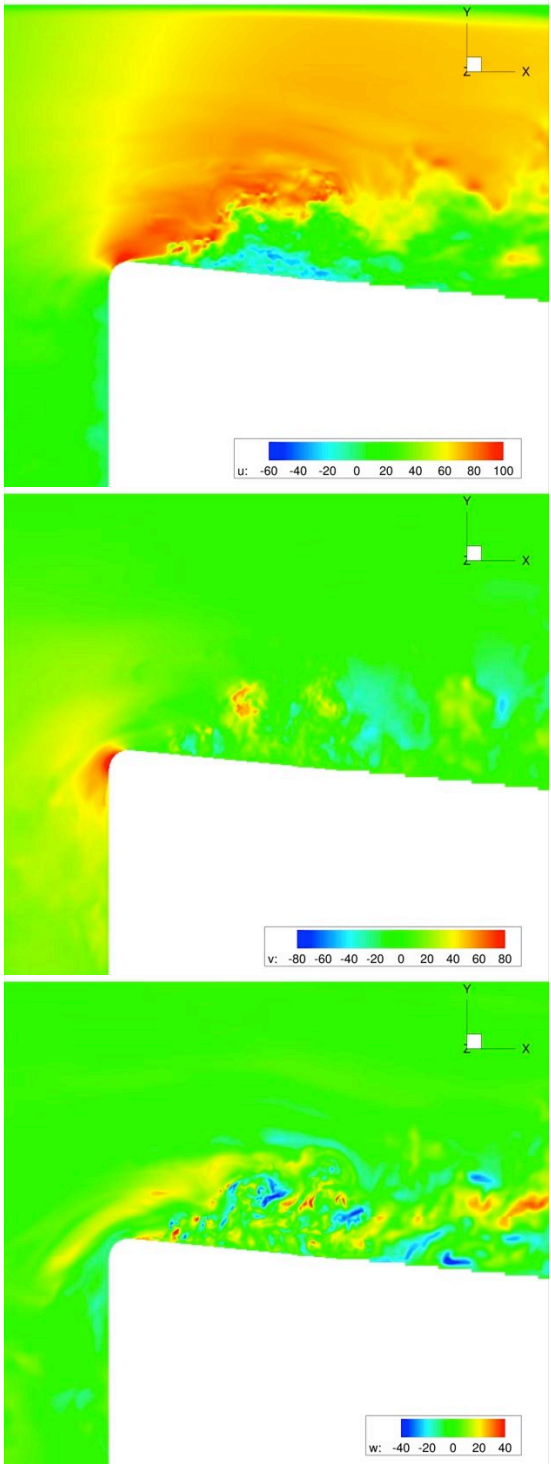

**Fig. 6:** Instantaneous velocity field in the center plane for LES of cavitating forward-facing step flow in a micro channel.

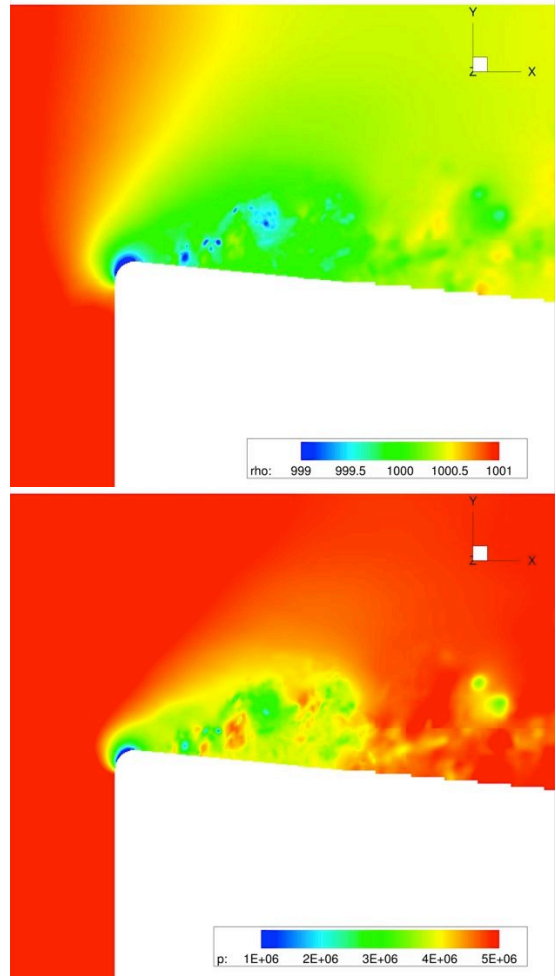

**Fig. 7:** Instantaneous density and pressure field in the center plane for LES of cavitating forward-facing step flow in a micro channel.

### ACKNOWLEDGMENTS

We would like to thank Christian Egerer for visualizing the computational results. Simulations were performed at the Leibniz-Rechenzentrum München (LRZ).

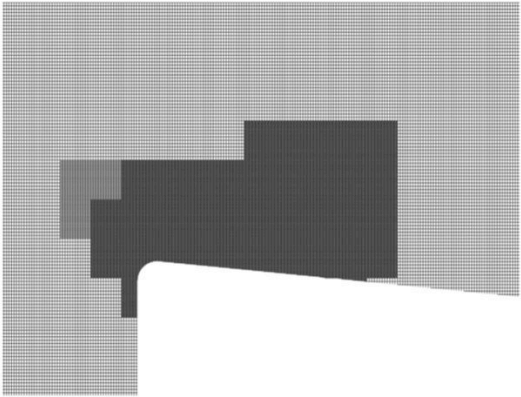
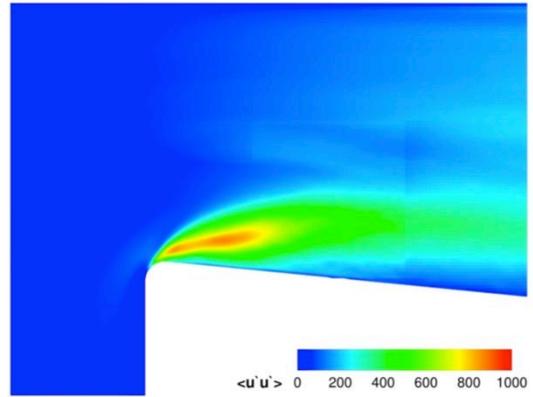
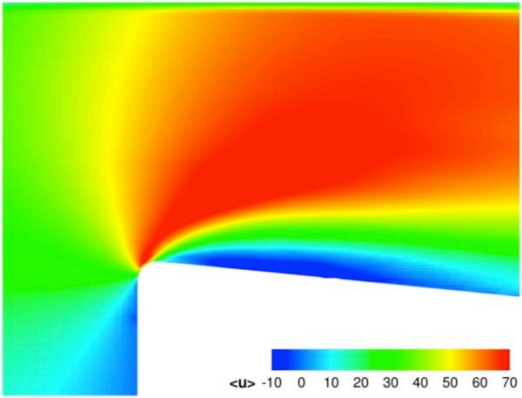
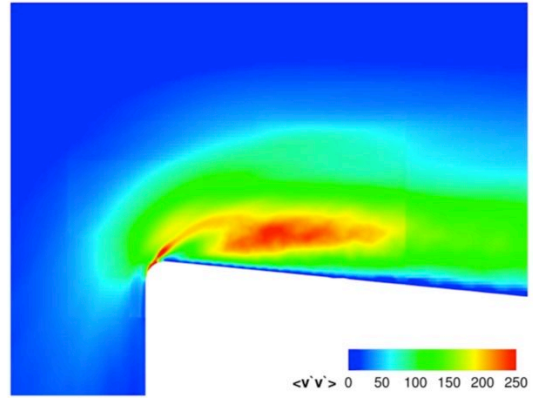
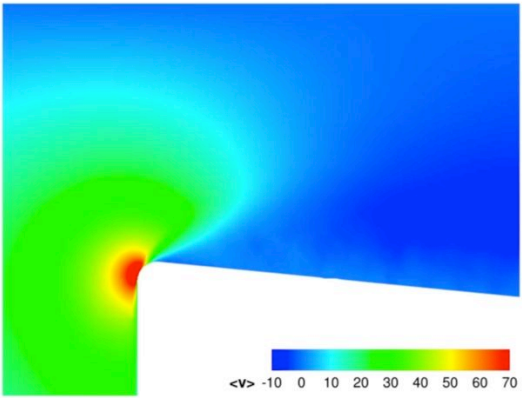
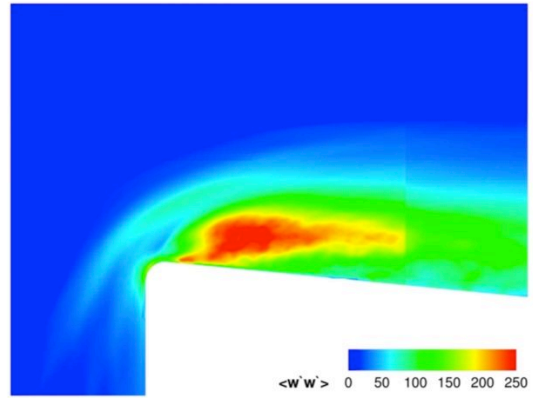
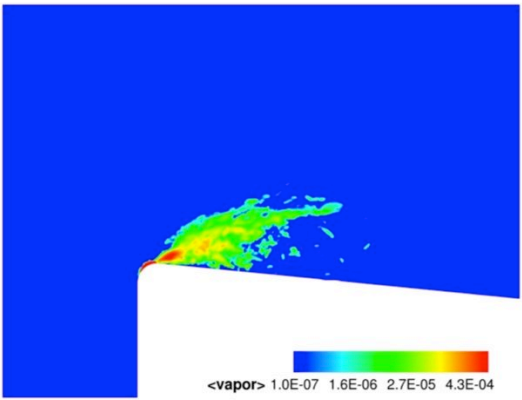
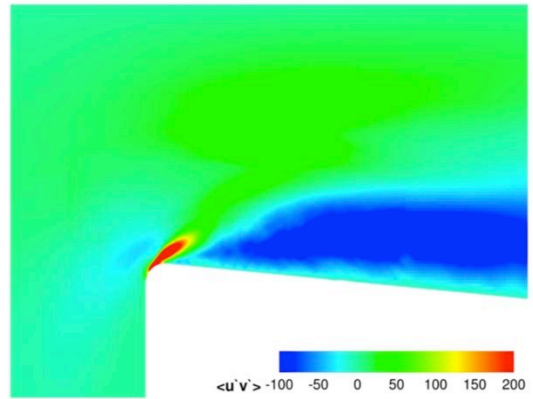

**Fig. 8:** Adaptive Cartesian computational grid in the interrogation region, mean velocity field and mean vapor volume fraction.

**Fig. 9:** Reynolds stresses.